\documentclass{birkjour}

\usepackage[utf8]{inputenc}
\usepackage[T1]{fontenc}
\usepackage{cite}
\usepackage{graphicx} 

\newtheorem{thm}{Theorem}

\theoremstyle{definition}
\newtheorem{defn}{Definition}
\theoremstyle{remark}

\begin{document}

\title[Quasicrystal problem]{Quasicrystal problem - on rigidity of\\
  non-periodic structures from statistical\\
  mechanics point of view} \author{Jacek Miękisz}
\address{University of Warsaw \\
  Institute of Applied Mathematics and Mechanics \\
  Banacha 2, 02-097 Warsaw, Poland} \email{miekisz@mimuw.edu.pl}



\begin{abstract}
  We present a brief history of quasicrystals and a short introduction
  to classical lattice-gas models of interacting particles.  We
  discuss stability of non-periodic tilings and one-dimensional
  sequences of symbols seen as ground states of some hamiltonians. We
  argue that some sort of homogeneity, the so-called Strict Boundary
  Condition, is necessary for stability of non-periodic ground states
  against small perturbations of interactions and thermal
  fluctuations.
\end{abstract}

\maketitle

\section{Introduction}

Since the discovery of quasicrystals by Dan Shechtman
\cite{shechtman}, one of the fundamental problems in condensed matter
physics is to understand their occurrence in microscopic models of
interacting particles. However, the history of investigating
non-periodic structures is much longer and it took place mainly in
mathematical circles. On the other side, physicists always assumed
that any matter at sufficiently low temperatures and high enough
pressure is crystalline that is periodic. Such a hypothesis, the
so-called Crystal Problem, was never proved but it was discussed and
investigated by some mathematical physicists
\cite{simon,radinreview1,radinreview2}.

Let us begin our short history of quasicrystals in 1900 when David
Hilbert presented his famous 23 problems.  The second part of the
18-th problem can be phrased in the following way: does there exist a
polyhedron which can cover the space but only in a non-periodic way?
More precisely, we have at our disposal an infinite number of copies
of a polyhedron and we can cover by them the three-dimensional space
such that all points of the space are covered and there are no
overlaps (with the exception of boundaries of polyhedra). Moreover,
each covering is non-periodic - they are not invariant under a
non-trivial translation.

In 1961, Hao Wang translated the Hilbert problem for domino players
\cite{wang}.  We have now a finite number of squares with colored
sides. Wang's conjecture went in the opposite direction: if we can
tile the plane with dominoes such that colors of neighboring sides are
the same, we can also tile it in a periodic way.  The first
counter-example was constructed in 1972 (or discovered if you wish) by
Richard Berger \cite{berger}.  He used over 20 thousand different
types of dominoes. In 1971, this number was reduced by Raphael
Robinson \cite{robinson} to 56 square-like tiles (tiles with notches
and dents) or $6$ if you allow rotations and reflections. Let us note
that in any tiling, the centers of squares form a regular periodic
lattice $Z^{2}$, non-periodic are assignments of tiles to lattice
sites.  In 1974, Roger Penrose constructed two polygons, famous kite
and dart (they are not square-like Wang tiles), which cover the plane
only in a non-periodic way \cite{penrose}, see also a recent
introductory text \cite{francesco}.

In 1981, Dan Shechtman observed a five-fold symmetry in diffraction
patterns of rapidly solidified aluminum transition metal alloys
\cite{shechtman}.  Such symmetry is forbidden in periodic
lattices. Quasicrystals were born. We have to mention here that
exactly the same diffraction pattern was observed before in Penrose
tilings with point masses located in their vertices.

In meantime, non-periodic bi-infinite sequences on $Z$ were
investigated in the framework of symbolic dynamical systems.  Examples
include Thue-Morse and Fibonacci sequences or in general Sturmian
systems \cite{queffelec}.
 
In all cases considered here, when one picks any non-periodic
configuration, construct an infinite orbit under translations and
close it in the product topology, one gets a compact set which
supports only one translation-invariant measure. It means that in any
such model, all non-periodic configurations look the same, one cannot
distinguish them locally, they have the same frequency of all local
patches.  More precisely, there exists a unique ergodic
translation-invariant probability measure concentrated on them.  It
can be seen as the ``average'' of point measures corresponding to
those configurations.

The main problem in our research is to find out how rigid are such
non-periodic structures. As a first step we look at relatively simple
toy hamiltonians for which they are their ground-state configurations,
that is configurations which minimize energy of the system of
interacting particles in the so-called lattice-gas models. Such models
have unique translation-invariant ground-state measures, called
non-periodic ground state, supported by ground-state configurations.
Then the fundamental question is how generic are such examples. We ask
whether any small perturbation of a given hamiltonian has the same
unique non-periodic ground state. If this is the case we say that it
is zero-temperature stable.

In positive temperatures, our systems of interacting particles undergo
stochastic fluctuations caused by random thermal motions.  The system
is low-temperature stable if non-periodic configurations survive such
thermal perturbations, that is a unique non-periodic ground state
gives rise to a low-temperature equilibrium - a non-periodic Gibbs
measure which is a small perturbation of the ground state.

We argue that some sort of homogeneity, the so-called Strict Boundary
Condition \cite{miesta, strictboundary} is necessary for stability of
non-periodic ground states.

If non-periodic ground-state configurations are zero and
low-temperature stable we may call a corresponding lattice-gas model a
microscopic model of a quasicrystal. For a comprehensive exposition of
mathematical models of quasicrystals we refer readers to the recent
book of Michael Baake and Uwe Grimm \cite{baake}.

\section{A short introduction to lattice-gas models}

This is a very concise introduction to lattice-gas models in
statistical mechanics. For an extensive presentation one may consult
the recent monograph available on-line \cite{intro} and also texts on
mathematical statistical mechanics \cite{israel,bricmont}.  In
lattice-gas models, particles, spins or other entities are assigned to
vertices of graphs, usually regular lattices but some other networks
like random graphs can also be considered. Here we will consider
$Z^{d}$, $d=1, 2$.  Let $S$ be a finite set of types of particles,
spins orientations or abstract symbols.  $\Omega = S^{Z^{d}}$ is the
space of all microscopic configurations of our system, that is
assignments of entities to lattice sites.  $\Omega$ equipped with the
product topology of discrete topologies on the set of one-site
configurations is a compact space.  For any finite
$\Lambda \subset Z^{d}$, $\Omega_{\Lambda} = S^{\Lambda}$ is the set
of finite configurations.  For $X \in \Omega$ (or $\Omega_{\Lambda}$)
and $i \in Z^{d}$, we denote by $X_{i} \in S$ a configuration at the
lattice site $i$.
 
Now we would like to introduce interactions between particles or
spins. They are described by a family of functions for all finite
$\Lambda$'s,
$$\Phi_{\Lambda}: \Omega_{\Lambda} \rightarrow R, \hspace {2mm} \Lambda \subset Z^{d}.$$
Then we introduce a finite-volume $V$ Hamiltonian,
$$H_{V} = \sum_{\Lambda \subset V} \Phi_{\Lambda}.$$
Usually we will consider two-body interactions, that is
$\Phi_{\Lambda} = 0$ if $\Lambda \neq \{i,j\}, i, j \in \Lambda.$ We
may also have one-site interactions, $\Phi_{i}$, interpreted as an
external magnetic field or a chemical potential.

At zero temperature, any classical system of interacting particles
should minimize its energy, it should be at a ground state.  Infinite
systems have in general infinite energies so we cannot define
ground-state configurations as those which minimize the
energy. Another possible definition of ground-state configurations as
those which minimize the energy density (energy per lattice site) is
not appropriate for the following reason. When one changes a
ground-state configuration at a finite region (a local excitation),
the energy density does not change so the excited configuration should
also be called a ground-state configuration which is an absurd.

We will introduce the following definition which is particularly
useful in systems without periodic ground-state configurations.
\begin{defn} \label{defn:1}
  Let $Y, X \in \Omega$. $Y$ is a {\bf local excitation} of $X$, denoted $Y \sim X$, \\
  if $|\{i, Y_{i} \neq X_{i}\}| < \infty$.
\end{defn}
\begin{defn} \label{defn:2} Let $Y \sim X$,  $H(Y|X) = \sum_{\Lambda} (\Phi_{\Lambda} (Y) - \Phi_{\Lambda} (X))$ \\
  is a {\bf relative Hamiltonian}.
\end{defn}
Observe that for finite-range interactions, the sum is finite. For
infinite-range interactions, to have a convergent series we have to
assume that interactions decay at infinity reasonably fast, we should
consider the so-called summable interactions.
\begin{defn}\label{defn:3}
  $X \in \Omega$ is a {\bf ground-state configuration} if for any
  local excitation $Y \sim X$,
  $$H(Y|X) \geq 0.$$
\end{defn}
It can be proved that for any finite-range (or summable) interactions
the set of ground-state configurations is non-empty and they minimize
the energy density. However, as we will see in the next section, there
are lattice-gas models without any periodic ground-state
configurations. Recall that a configuration is periodic if it is
invariant under some lattice translation.

Here we will consider models without periodic ground-state
configurations with the following properties. In each model, all its
non-periodic ground-state configurations look the same, they cannot be
distinguished locally. More precisely, any local patch (configuration
on a finite subset of $Z^{d}$) which is present in one ground-state
configuration appears with the same frequency in all of them. Let us
describe this property more precisely.

Let $a \in Z^{d}$ and $T_{a}$ be a translation operator, that is if
$X \in \Omega$, then $(T_{a}X)_{i} = X_{i-a}$.  Let $X$ be any
ground-state configuration of our model. We construct a d-dimensional
orbit of $X$, $\{T_{a}X, a \in Z^{d}\}$. Its closure does not depend
on $X$, we denote it by $G$. It is an uncountable compact subset of
$\Omega$, the set of all ground-state configurations (without fault
lines separating two ground-state configurations). In all our models,
$G$ supports a unique translation-invariant probability measure called
the ground state, denoted by $\mu_{G}.$ It can be constructed as the
limit of "D ``Dirac combs''. Namely,
$$\mu_{G} = \lim_{\Lambda\rightarrow Z^{d}} \frac{1}{|\Lambda|}\sum_{a \in \Lambda} \delta_{T_{a}X},$$
where $\delta_{Y}$ is the measure assigning the probability $1$ to
$Y$, for the infinite-volume limit, $\Lambda\rightarrow Z^{d}$, we may
use hypersquares of increasing sizes, the limit of sequence of
measures is taken in the weak* topology.

The triple $(G, T, \mu_{G})$ is called a symbolic dynamical system,
$\mu_{G}$ is a uniquely ergodic measure with respect to $T$ - a
homeomorphism from $G$ to~$G$.

There is one important property of non-periodic ground-state
configurations which goes beyond unique ergodicity, the so-called
Strict Boundary Condition \cite{miesta,strictboundary}, also called
rapid convergence to equilibrium of frequencies
\cite{peyriere,gambaudo}, for a related notion of hyperuniformity see
\cite{torquato1,torquato2,maher} and the Appendix. It is defined for
non-frustrated Hamiltionians.
\begin{defn}\label{defn:4}
  Interactions $\Phi$ are {\bf non-frustrated} if there exists
  $X \in \Omega$ such that for every $\Lambda$,
  $\Phi_{\Lambda}(X) = \min_{Y \in \Omega}\Phi_{\Lambda}(Y)$.
\end{defn}
For finite-range interactions we may assume that minima are equal
to~$0$ and all other interactions are equal to $1$.  For
infinite-range interactions, energies should converge to zero
sufficiently fast with distance between particles.  Let us note that
any configuration for which all interactions attain zero values is a
ground-state configuration.

Let $ar \in S^{\Lambda_{ar}}$ be a local patch.  Let $X$ be a
ground-state configuration and $Y \sim X$, $H(Y|X) = B(Y)$ is the
number of broken bonds, that is the number of $\Lambda$ such that
$\Phi_{\Lambda}(Y) = 1$ and $n_{ar}(Y|X)$ is the difference of the
number of appearances of $ar$ in $Y$ and in $X$.
\begin{defn}\label{defn:5}
  A classical lattice-gas model satisfies the {\bf Strict Boundary
    Condition} if for any local patch $ar$, there exists $C_{ar}$ such
  that for every $Y \sim X$,
  $$|n_{ar}(Y|X)| < C_{ar}B(Y).$$
\end{defn}  
We may also define the Strict Boundary Condition for tilings,
non-periodic configurations in general, without any reference to
interactions, see the Appendix.

We would like to see whether unique ground-state measures of our
hamiltonians are stable against small perturbations of interactions
and thermal motions of particles.
\begin{defn}\label{defn:6}
  A ground-state measure of a hamiltonian $H$ is zero-temperature
  stable if it is a ground state of any perturbed hamiltonian
  $H + \epsilon H'$ for a finite-range $H'$ and a sufficiently small
  $\epsilon$.
\end{defn}
We have the following theorem connecting the Strict Boundary Condition
with the zero-temperature stability \cite{miesta}.
\begin{thm}\label{thm:1} A unique ground-state measure of a
  finite-range Hamiltonian is stable against small perturbations of
  interactions of the range smaller than~$r$ if and only if the Strict
  Boundary Condition is satisfied for patches of a diameter smaller
  than $r$.
\end{thm}

At positive temperatures, we would like to see if a ground-state
measure survives thermal motions and it gives rise to a Gibbs state, a
measure which is supported by small perturbations of ground-state
configurations.  A finite-volume $V$ Gibbs state (called a grand
canonical ensemble in statistical physics) describes probabilities of
microscopic configurations at an equilibrium. It is given by the
following formula,
$$\rho_{V, H_{V}, T} = \frac{e^{-\beta H_{V}}}
{\sum_{Y \in \Omega_{V}} e^{-\beta H_{V}(Y)}},$$ where $\beta = 1/T$.

In an analogous way we define a finite-volume Gibbs state
$\rho^{X}_{V, H_{V}, T}$ when in the above definition we put a
ground-state configuration $X$ outside $V$ and allow interactions
between particles inside and outside $V$.

An infinite-volume Gibbs measure $\rho^{X}_{H, T}$ is defined as a
limit of $\rho^{X}_{V, H_{V}, T}.$
\begin{defn}\label{defn:7} A ground-state configuration $X$ is {\bf
    low-temperature stable} if at sufficiently small temperatures
  there exists a Gibbs measure $\rho^{X}$ which is a small
  perturbation of $\delta_{X}$.
\end{defn}
The general theory of Gibbs measures as perturbation of ground states
was developed by Pirogov and Sinai in \cite{ps1,ps2}. Some results
concerning stability of non-periodic ground states are presented in
the following sections.

\section{Two-dimensional models based on the Robinson's tilings}

Let us recall that in any Robinson's tiling, centers of square-like
tiles (there are 56 of them, we consider a rotated or reflected tile
as a different one) form a regular periodic lattice $Z^{2}$.  Denote
by $R \subset \Omega = \{1, ..., 56\}^{Z^{2}}$ the set of all
Robinson's tilings.  $(R, T,\mu_{R})$ is a uniquely ergodic system.

Now we will construct a lattice-gas model based on Robinson's
tilings~\cite{radin}. First we identify tiles with particles,
therefore there are $56$ types of particles. If two tiles do not
match, then we set the interaction energy between particles
corresponding to them to $1$, otherwise the interaction energy is
$0$. Ground-state configurations of our lattice-gas model are given by
infinite tilings, all matching rules are satisfied so the interaction
energy between any pair of particles is equal to $0$.

It can be shown that lattice-gas model based on Robinson's tilings
does not satisfy the Strict Boundary Condition \cite{jaradin1}.  There
are arbitrary long sequences of tiles, called arrows, pointing in one
direction such that if one changes them to tiles with arrows pointing
in the opposite direction, there are broken bonds (pairs of particles
with the positive interaction energy) only at the end of
sequences. One may therefore introduce an arbitrary small chemical
potential favoring tiles with an arrow in one direction. We have the
following theorem \cite{jaradin1}.
\begin{thm}\label{thm:2}
  Robinson’s ground state is not stable against an arbitrarily small
  chemical potential favoring one type of particles.
\end{thm}
There are some partial results concerning low-temperature stability
\cite{manyphases}, the best one is based on a modified Robinson'
tilings \cite{sequenceoftemp}.
\begin{thm}\label{thm:23} There is a decreasing sequence of
  temperatures $T_{n}$ such that if $T<T_{n}$, then there exists a
  Gibbs state with a period at least $6^{n}$ in both directions.
\end{thm}
We put forward a hypothesis that homogeneity property present in the
Strict Boundary Condition is necessary for the low-temperature
stability of non-periodic ground states of finite-range interactions.

\section{One-dimensional models based on dynamical systems}

It is known that any one-dimensional lattice-gas models with
finite-range interactions has at least one periodic ground-state
configuration \cite{bundangnenciu,schulradin,thirdlaw}. Therefore to
force non-periodicity we have to consider models with infinite-range
potentials. Below we review constructions of one-dimensional
hamiltonians without periodic ground-state configurations, based on
one-dimensional uniquely ergodic symbolic dynamical systems,
Thue-Morse \cite{tmhamiltonian} and Sturmian ones \cite{ahj}.

\subsection{Thue-Morse system}

First we construct Thue-Morse sequences. We put $1$ at the origin and
perform successively the substitution:
$+ \rightarrow +-, - \rightarrow -+$. In this way we get a one-sided
sequence $+--+-++--++-+--+ ...$, $\{X_ {TM }(i)\}, i \geq 0$. We
define $X_{TM} \in \Omega^{Z} = \{0, 1\}^{Z}$
by setting $X_{TM}(i) = X_{TM} (-i-1 )$ for $i<0$. Let $G_{TM}$ be the
closure (in the product topology of discrete topologies on $\{-,+\}$)
of the orbit of $X_{TM}$ by the translation operator $T$, i.e.,
$G_{TM} = \{T^{n} (X_{TM}), n \geq 0\}^{cl}.$ It can be shown
\cite{keane} that $G_{TM}$ supports exactly one translation-invariant
probability measure $\mu_{TM}$ on $\Omega$.

Let us identify now $+$ with $+1$ and $-$ with $-1$.  It was shown in
\cite{tmhamiltonian} that $\mu_{TM}$ is the only ground state of the
following exponentially decaying four-spin interactions,
\begin{equation}
  H_{TM} = \sum_{r=0}^{\infty} \sum_{p=0}^{\infty} H_{r,p},
\end{equation}
where
\begin{equation}
  H_{r,p} = \sum_{i \in Z} J(r,p) (\sigma_{i} + \sigma_{i+2^{r}})^{2} (\sigma_{i+(2p+1)2^{r}} + \sigma_{i+(2p+2)2^{r}})^{2}
\end{equation}
and $\sigma_{i}(X)= X_{i} \in \{+1, -1\}$.  It was proved that the
Thue-Morse ground state is unstable \cite{nonstability}.
\begin{thm}\label{thm:4} The Thue-Morse ground state $\mu_{TM}$ is
  unstable against an arbitrarily small chemical potential which
  favors the presence of molecules consisting of two up or down
  neighboring spins.
\end{thm}

\subsection{Sturmian systems}

Here we will consider configurations of two symbols, $0$ and $1$, on
$Z$.

We will identify the circle $C$ with $R/Z$ and consider an irrational
rotation by $\varphi$ (which is given by translation on $R/Z$ by
$\varphi \; mod \; 1$) and let $P=[0,\varphi)$.
\begin{defn}\label{def:8} Given an irrational $\varphi\in C$ we say
  that $X\in\{0,1\}^Z$ is generated by $\varphi$ if it is of the
  following form, $X_{n}=0$ if $n \varphi \in P$ and $1$ otherwise.
\end{defn}
We call such $X$ a Sturmian sequence corresponding to $\varphi$.  Let
$G_{St}$ be the closure of the orbit of $X$ by translations.  It can
be shown that $G_{St}$ supports exactly one translation-invariant
probability measure $\rho_{St}$ on $\Omega$.

We will consider only rotations by badly approximable numbers.
\begin{defn}\label{def:9}
  We say that a number $\varphi$ is \textbf{badly approximable} if
  there exists $c>0$ such that
$$\left | \varphi - \frac{p}{q} \right | > \frac{c}{q^2} $$
for all rationals $\frac{p}{q}$.
\end{defn}
Sturmian sequences can be characterized by absence of certain finite
patches \cite{ahj}.
\begin{thm}\label{thm:5}
  Let $\varphi\in (\frac{1}{2},1)$ be irrational. Then there exist a
  natural number $m$ and a set $F\subseteq N$ of forbidden distances
  such that Sturmian sequences generated by $\varphi$ are uniquely
  determined by the absence of the following patterns: $m$ consecutive
  0's and two 1's separated by a distance from $F$.
\end{thm}
For $\varphi = \frac{2}{1+ \sqrt{5}}$ we get Fibonacci sequences
produced by the substitution rule $0 \rightarrow 01, 1 \rightarrow 0.$

Now we construct hamiltonians having Sturmian measures as unique
ground-states. We assign positive energies to forbidden patterns and
zero otherwise and have the following theorem \cite{ahj}.
\begin{thm}\label{thm:5}
  For every Sturmian system there exist non-frustrated, arbitrarily
  fast decaying, two-body interactions (augmented by a finite-range
  non-frustrated interaction penalize $m$ successive 0's) for which
  the unique ergodic translation-invariant ground-state measure is the
  ergodic measure of the Sturmian system.
\end{thm}
We have a non-stability result.
\begin{thm}\label{thm:6}
  Assume that $\varphi \in (0,1)$ is badly approximable and the
  interaction energy decays as $1/r^{\alpha}$ with $\alpha>3$.  Then
  Sturmian ground-state configurations generated by $\varphi$ are
  unstable against arbitrarily small chemical potential of particles.
\end{thm}

Let us mention that one-dimensional two-body interactions having
Sturmian ground states were presented in
\cite{bakbruinsma,aubry2,aubry3,jedmiek1,jedmiek2}.  Hamiltonians in
these papers consist of two-body repelling interactions between
particles and a chemical potential favoring them.  Such interactions
are obviously frustrated. Ground states of these models form Cantor
sets called devil's staircases.  Non-periodicity is present only for
certain values (of measure zero) of chemical potentials - an
arbitrarily small change of a chemical potential destroys a
non-periodic ground state.

We do not know any examples of one-dimensional hamiltonians with
unique non-periodic ground states stable against small perturbations
of interactions. However, non-periodic Gibbs states for slowly
decaying (but summable) interactions were constructed in
\cite{enterja1,zahradnik}.

\section{Open problems}

\begin{enumerate}
\item Does there exist a uniquely ergodic non-periodic tiling system
  which satisfies the Strict Boundary Condition?

\item Does there exists a a classical lattice-gas model with
  finite-range interactions and with a unique non-periodic
  ground-state measure which is stable with respect to small
  perturbations of interactions?

\item Are Sturmian ground states stable with respect to perturbations
  of small finite-range interactions?

\item Does there exist a lattice-gas model with finite-range
  interactions and with a unique non-periodic ground-state measure
  which gives rise to a low-temperature non-periodic Gibbs state?
\end{enumerate}
For some other open problems in mathematical models of quasicrystals
we refer readers to \cite{sadun}.

\section*{Appendix - Strict Boundary Condition in uniquely ergodic
  systems}

First we discuss non-periodic tilings of the plane by square-like Wang
tiles, like Robinson' tilings. In any such tiling, centers of tiles
form the lattice $Z^{2}$.  Denote by
$G \subset \Omega = \{1, ..., n\}^{Z^{2}}$ the set of all tilings with
$n$ different types of tiles.  We consider tilings such that
$(G, T,\mu_{G})$ is a uniquely ergodic system, where $G$ consists of
only non-periodic tilings and $\mu_{G}$ is a unique probability
translation-invariant measure with $G$ as its support.  Therefore all
tilings have the same density of any given local patch
$ar \in \{1,...,n\}^{\Lambda_{ar}}$, denote it by $\omega_{ar}$.

Let $X \in \{1,...,n\}^{\Lambda}$ be a local tiling on a finite
$\Lambda \subset Z^{2}$, all matching rules in $\Lambda$ are
satisfied, however it might not be extendable to an infinite tiling,
that is an element of $G$.  Let $|\Lambda|$ be the number of sites in
$\Lambda$ and $P(\Lambda)$ the length of its boundary.
$$n_{ar}^{\Lambda}(X) = |V \subset  \Lambda, V = T_{b}\Lambda_{ar}, b \in {\bf Z}^{2}; X_{V} = ar|$$
is the number of occurrences of the patch $ar$ in $\Lambda.$
\begin{defn} \label{defn:10} $G$ satisfies the {\bf Strict Boundary
    Condition} if for any local patch $ar$ there exists $C_{ar}$ such
  that for any local tiling $X$ of a finite set $\Lambda$ we have

$$|n_{ar}^{\Lambda}(X) - \omega_{ar}|\Lambda|| < C_{ar}P(\Lambda).$$
\end{defn}
Such property is also called the rapid convergence to equilibrium of
frequency of patches \cite{peyriere,gambaudo}.

This is the requirement that the number of any finite patch in a
finite lattice subset can fluctuate at most proportional to its
boundary.  One can easily generalize this definition to tilings with
not necessarily square-like tiles, like Penrose tilings.

In meantime some other related concepts were introduced: Bounded
Fluctuations \cite{aizenman} and Hyperuniformity
\cite{torquato1,torquato2,maher}.  In both cases, the authors demand
that the variance of the number of particles in a bounded domain
should grow slower than its boundary in the limit of the infinite
volume. In comparison, the Strict Boundary Condition concerns
many-particle patches at all length scales, not only in the
infinite-volume limit.  Moreover, it is a deterministic concept, not a
stochastic one and it refers directly to the minimization of
interactions.

As it was described in Section 3, with any uniquely ergodic tiling
system we can associate a classical lattice-gas model with a unique
non-periodic ground-state measure. We can reformulate now the Strict
Boundary Property for ground-state configurations of non-frustrated
hamiltonians with finite-range interactions (not necessarily based on
tilings).

Let $X \in \{1,...,n\}^\Lambda$ be a local ground-state configuration,
that is a configuration on $\Lambda$ which minimizes all interactions.
\begin{defn} \label{defn:10}\noindent $G$ satisfies the {\bf Strict
    Boundary Condition} for local ground-state configurations if for
  any local patch $ar$ there exists a constant $C_{ar}$ such that for
  any local tiling $X$ on $\Lambda$,
$$|n_{ar}^{\Lambda}(X) - \omega_{ar}|\Lambda|| < C_{ar}P(\Lambda).$$
\end{defn}
One can prove that Definitions 5 and 10 of the Strict Boundary
Condition are equivalent \cite{strictboundary}.

One may also formulate an analogous condition for a configuration on
$Z$.
\begin{defn} \label{defn:10} $X \in \{1,...,n\}^{Z}$ satisfies the
  {\bf Strict Boundary Condition} if for any local patch $ar$ there
  exists a constant $C_{ar}$ such that for any $A$ consisting of $L$
  consecutive lattice sites
$$|n_{ar}^{A}(X) - \omega_{ar}L| < C_{ar}.$$
\end{defn}
There are other notions of homogeneity of one-dimensional
configurations: most-homogeneous configurations discussed in
\cite{bakbruinsma,aubry2,aubry3} and balanced property. It was shown
in \cite{ahj} that these three notions of homogeneity are equivalent
and that Thue-Morse sequences do not satisfy the Strict Boundary
Condition as opposed to the Sturmian systems.

\subsection*{Acknowledgments} I would like to thank the National
Science Centre (Poland) for a financial support under Grant
No. 2016/22/M/ST1/00536.

\end{document}